\documentclass[runningheads,a4paper]{llncs}

\usepackage{amssymb}
\usepackage{graphicx}

\usepackage{url}
\urldef{\mailsa}\path|armin.balalaie@gmail.com|
\urldef{\mailsb}\path|p.jamshidi@imperial.ac.uk|
\urldef{\mailsc}\path|, heydarnoori@sharif.edu|
\newcommand{\keywords}[1]{\par\addvspace\baselineskip
\noindent\keywordname\enspace\ignorespaces#1}

\usepackage[colorinlistoftodos,prependcaption,textsize=tiny]{todonotes}
\newcommand{\unsure}[2][1=]{\todo[linecolor=red,backgroundcolor=red!25,bordercolor=red]}
\newcommand{\change}[1][1=]{\todo[linecolor=blue,backgroundcolor=blue!25,bordercolor=blue]}
\newcommand{\info}[2][1=]{\todo[linecolor=OliveGreen,backgroundcolor=OliveGreen!25,bordercolor=OliveGreen]}
\newcommand{\improvement}[2][1=]{\todo[linecolor=Plum,backgroundcolor=Plum!25,bordercolor=Plum]}
\newcommand{\thiswillnotshow}[2][1=]{\todo[disable]}

 \newcommand{\comment}[1]{}

\newcommand{\secref}[1]{Section~\ref{sec:#1}}

\newcommand{\figref}[1]{Figure~\ref{fig:#1}}

\usepackage[caption=false]{subfig}

\begin{document}

\mainmatter  

\title{Migrating to Cloud-Native Architectures Using Microservices: An Experience Report}

\titlerunning{Migrating to Cloud-Native Architectures Using Microservices}

\author{Armin Balalaie\inst{1}\and Abbas Heydarnoori\inst{1}\and Pooyan Jamshidi\inst{2}}
\authorrunning{Migrating to Cloud-Native Architectures Using Microservices}

\institute{Sharif University of Technology, Tehran, Iran\\
\mailsa
\mailsc\\
\and Department of Computing, Imperial College London, UK\\
\mailsb\\}

\maketitle

\begin{abstract}
Migration to the cloud has been a popular topic in industry and academia in recent years. Despite many benefits that the cloud presents, such as high availability and scalability, most of the on-premise application architectures are not ready to fully exploit the benefits of this environment, and adapting them to this environment is a non-trivial task. Microservices have appeared recently as novel architectural styles that are native to the cloud. These cloud-native architectures can facilitate migrating on-premise architectures to fully benefit from the cloud environments because non-functional attributes, like scalability, are inherent in this style. The existing approaches on cloud migration does not mostly consider cloud-native architectures as their first-class citizens. As a result, the final product may not meet its primary drivers for migration. In this paper, we intend to report our experience and lessons learned in an ongoing project on migrating a monolithic on-premise software architecture to microservices. We concluded that microservices is not a one-fit-all solution as it introduces new complexities to the system, and many factors, such as distribution complexities, should be considered before adopting this style. However, if adopted in a context that needs high flexibility in terms of scalability and availability, it can deliver its promised benefits.

\keywords{Cloud Migration, Microservices, Cloud-native Architectures, Software Modernization}
\end{abstract}

\section{Introduction}
In recent years, with the emergence of cloud computing and its promises, many companies from large to small and medium sizes are considering cloud as a target platform for migration~\cite{jamshidi2013migrationreview}. Despite motivations for migrating to the cloud, most of the applications could not benefit from the cloud environment as long as their main intention is to simply dump the existing legacy architecture to a virtualized environment and call it a cloud application.

One of the main characteristics of the cloud environment is that failures can happen at any time, and the applications in this environment should be designed in a way that they can resist such uncertainties. Furthermore, application \emph{scalability} would not be possible without a scalable architecture. Cloud-native architectures like \emph{microservices} are the ones that have these characteristics, i.e., availability and scalability, in their nature and can facilitate migrating on-premise architectures to fully benefit from the cloud environments.

Microservices is a novel architectural style that has been proposed to overcome the shortcomings of a \emph{monolithic architecture}~\cite{richardson2014micropatterns} in which the application logic is within one deployable unit. For small systems, the monolithic architecture could be the most appropriate solution and could become highly available and scalable using simple load balancing mechanisms. However, as the size of the system starts growing, problems like difficulties in understanding the code, increased deployment time, scalability for data-intensive loads, and a long-term commitment to a technology stack would start to appear~\cite{richardson2014micropatterns}. This is where microservices come to help by providing small services that are easy to understand, could be deployed and scaled independently, and could have different technology stacks.

Most of the current approaches on cloud migration are focused on automated migration via applying model-driven approaches~\cite{ardagna2012modaclouds,bergmayr2013migrating}, and reusing of knowledge by migration patterns~\cite{fehling2013service,jamshidicloud,mendonca2014architectural} without having cloud-native architectures as their first-class citizens. Furthermore, microservices is a new concept and thus, only a few technical reports can be found about using them in the literature~\cite{borsje2014microbuild,caldao2014microbuild}.

Migrating an application's architecture to microservices brings in many complexities that make this migration a non-trivial task. In this paper, we report our experience on an ongoing project in PegahTech Co.\footnote{http://www.pegahtech.ir}, on migrating an on-premise application named \emph{SSaaS} to microservices architecture. Although the migration steps that we describe in this paper are specific to our project, the necessity of performing these migration activities could be generalized to other projects as well. Furthermore, we summarize some of the challenges we faced and the lessons learned during this project.

The rest of this paper is organized as follows: \secref{background} briefly explains the background behind the microservices architecture. \secref{current_architecture} describes the architecture of SSaaS before its migration to microservices. The target architecture to which we migrated SSaaS is described in \secref{target_architecture}. \secref{migration_steps} then discusses our migration plan and the steps that we followed in our migration project. Next, \secref{lessons_learned} summarizes the lessons learned in this project. Finally, \secref{conclusion} concludes the paper.

\section{Background} \label{sec:background}
Microservices is a new trend that binds closely to some other new concepts like \emph{Continuous Delivery} and \emph{DevOps}. In this section, we first explain these concepts followed by the background on microservices architecture.

\subsection{Continuous Delivery and DevOps}
Continuous Delivery~\cite{humble2010continuous} is a software development discipline that enables on demand deployment of a software to any environment. With Continuous Delivery, the software delivery life cycle will be automated as much as possible. It leverages techniques like \emph{Continuous Integration} and \emph{Continuous Deployment} and embraces \emph{DevOps}. The DevOps is a culture that emphasizes the collaboration between developers and operations teams from the beginning of every project in order to reduce time to market and bring agility to all the phases of the software development life cycle. By adapting microservices, the number of services will be increased. Consequently, we need a mechanism for automating the delivery process.

\subsection{Microservices}
Microservices is a new architectural style~\cite{fowler2014microservices} that aims to realize software systems as a package of small services, each deployable on a different platform, and running in its own process while communicating through lightweight mechanisms like RESTFull APIs. In this setting, each service is a business capability which can utilize various programming languages and data stores. A system has a microservices architecture when that system is composed of several services without any centralized control~\cite{robert2014cleanmicroservices}. Resilience to failure is another characteristic of microservices as every request in this new setting will be translated to several service calls through the system. The Continuous Delivery and DevOps are also needed to be agile in terms of development and deployment~\cite{fowler2014microservices}.

To have a fully functional microservices architecture and to take advantage of all of its benefits, the following components have to be utilized. Most of these components address the complexities of distributing the business logic among the services:

\begin{itemize}
  \item \emph{Configuration Server:} It is one of the principles of Continuous Delivery to decouple source code from its configuration. It enables us to change the configuration of our application without redeploying the code. As a microservices architecture have so many services, and their re-deployment is going to be costly, it is better to have a configuration server so that the services could fetch their corresponding configurations.

  \item \emph{Service Discovery:} In a microservices architecture, there exist several services that each of them might have many instances in order to scale themselves to the underlying load. Thus, keeping track of the deployed services, and their exact address and port number is a cumbersome task. The solution is to use a Service Discovery component in order to get the available instances of each service.

  \item \emph{Load Balancer:} In order to be scalable, an application should be able to distribute the load on an individual service among its many instances. This is the duty of a Load Balancer, and in this case, it should get available instances from the Service Discovery component.

  \item \emph{Circuit Breaker:} Fault tolerance should be embedded in every cloud-native application, and it makes more sense in a microservices architecture where lots of dependent services are working together. Failure in each of this services may result in the failure of the whole system. Leveraging patterns like Circuit Breaker~\cite{nygard2007release} can mitigate the corresponding loss to the lowest level.

  \item \emph{Edge Server:} The Edge Server is an implementation of the \emph{API Gateway} pattern~\cite{richardson2014micropatterns} and a wall for exposing external APIs to the public. All the traffic from outside should be routed to internal services through this server. In this way, the clients would not be affected if the internal structures of system's services have changed afterwards.
\end{itemize}


\section{The Architecture of SSaaS Before the Migration} \label{sec:current_architecture}
The SSaaS (Server Side as a Service) application was initially started at PegahTech Co. to be a service that provides mobile application developers a facility for doing the server side programming part of their applications without knowing any server side languages. The PegahTech Co. envisions SSaaS as a service that could be scaled to millions of users. The first functionality of SSaaS was a RDBMS as a Service. Developers could define their database schema in the SSaaS website, and the SSaaS service provides them an SDK for their desired target platform (e.g., Android or iOS). Afterwards, the developers can only code in their desired platforms using their domain objects, and the objects would make some service calls on their behalf in order to fulfill their requests. As time goes on, new services are being added to SSaaS like Chat as a Service, Indexing as a Service, NoSQL as a Service, and so on.


SSaaS is written in Java using the Spring framework. The underlying RDBMS is an Oracle 11g. Maven is used for fetching dependencies and building the project. All of the services were in a Git repository, and the modules feature of Maven was used to build different services. At the time of writing this paper, there were no test cases for this project. The deployment of services in development machines was done using the Maven's Jetty plugin. However, the deployment to the production machine was a manual task that had many disadvantages~\cite{humble2010continuous}.

\begin{figure}[t]
  \centering
  \includegraphics[width=150px]{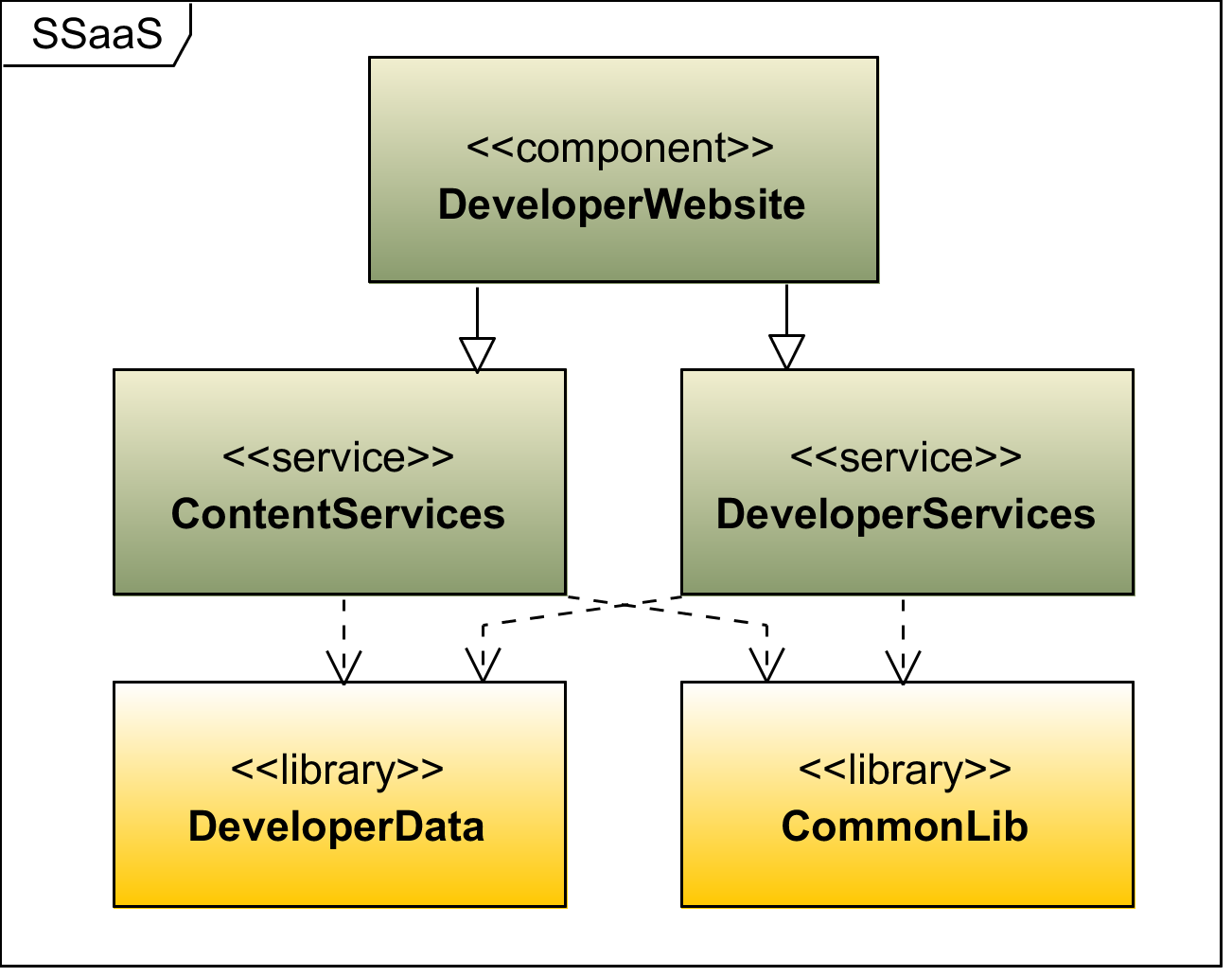}
  \caption{The architecture of SSaaS before the migration}\label{fig:current_architecture}
\end{figure}

In \figref{current_architecture}, solid arrows and dashed arrows respectively illustrate service calls direction and library dependencies. \figref{current_architecture} also indicates that SSaaS consisted of the following five components before the migration:

\begin{itemize}
\item \emph{CommonLib:} This is a place for putting shared functionalities, like utility classes, that are going to be used by the rest of the system.


\item \emph{DeveloperData:} This holds the information of developers who are using the SSaaS service and their domain model metadata entities that are shared between the DeveloperServices and the ContentServices components.


\item \emph{DeveloperServices:} This is where the services related to managing the domain model of developers' projects reside in. Using these services, developers could add new models, edit existing ones, and so on.

\item \emph{ContentServices:} This holds the services that the target SDK is using in order to perform the CRUD operations on the model's objects.

\item \emph{DeveloperWebsite:} This is an application written in HTML and JQuery and acts as a dashboard for developers. For this purpose, it leverages the DeveloperServices component.
\end{itemize}

\subsection{Why Did We Plan to Migrate towards the Microservices?}
What motivated us to perform a migration to a microservices architecture was a problem raised with a requirement for a Chat as a Service. To implement this requirement, we chose \emph{ejabberd}\footnote{\url{https://www.ejabberd.im/}} due to its known built-in scalability and its ability to run on clusters. To this end, we wrote a script in \emph{python} that enabled ejabberd to perform authentications using our system. After preparing everything, the big issue in our service was the \emph{on demand} capability, otherwise our service was useless. In the following, we discuss the reasons that motivated us to choose the microservices architecture:


\emph{The need for reusability:} To address the above issue, we started to automate the process of setting up a chat service. One of these steps was to set up a database for each user. We were hoping that this was also a step in creating RDBMS projects that we can reuse. After investigating the RDBMS service creation process, we recognized that there was not anything to satisfy our new requirement. To clarify further, there was a pool of servers in place. Each of these servers had an instance of the Oracle DBMS installed and an instance of DeveloperServices running. During the creation of a RDBMS project, a server was selected randomly and related users and tablespaces were created in the Oracle server. The mentioned design had several issues since it was just designed to fulfill the RDBMS service needs, and it was tightly coupled to the Oracle server. Nevertheless, we needed MySQL database for ejabberd and we should add this functionality to the system. After struggling a bit with the system, we recognized that we were just revamping the current bad design. What we needed was a database reservation system that both of our services could make use of. Thinking more generally, we needed a backing resources reservation system. This was the first step towards making cohesive services that can be reused by other parts of the system.

\emph{The need for decentralized data governance:} Another problem was that every time anyone wanted to add some metadata about different services, they were added to the DeveloperData. In other words, it was kind of an integration point among the services. It was not a good habit because services were independent units that were only sharing their contracts with other parts of the system. Consequently, another step was to re-architect the system so that any services could govern its own metadata and data by themselves.

\emph{The need for automated deployment:} As the number of services was growing, another problem was to automate the deployment process and to decouple the build life cycle of each service from other services as much as possible. This can happen using the Configuration Server and the Continuous Delivery components.

\emph{The need for built-in scalability:} As mentioned before, the vision of SSaaS is to serve millions of users. By increasing the number of services, we needed a new approach for handling this kind of scalability because scaling services individually needs a lot of work and can be error-prone. Therefore, to handle this problem, our solution was to locate service instances dynamically through the Service Discovery component and balancing the load among them using the internal Load Balancer component.

To summarize, new requirements pushed us to introduce new services, and new services brought in new non-functional requirements as mentioned above. Hence, we got advantage of microservices to satisfy these new requirements.

\section{The Target Architecture of SSaaS After the Migration} \label{sec:target_architecture}
\begin{figure}[t]
  \centering
  \includegraphics[width=250px]{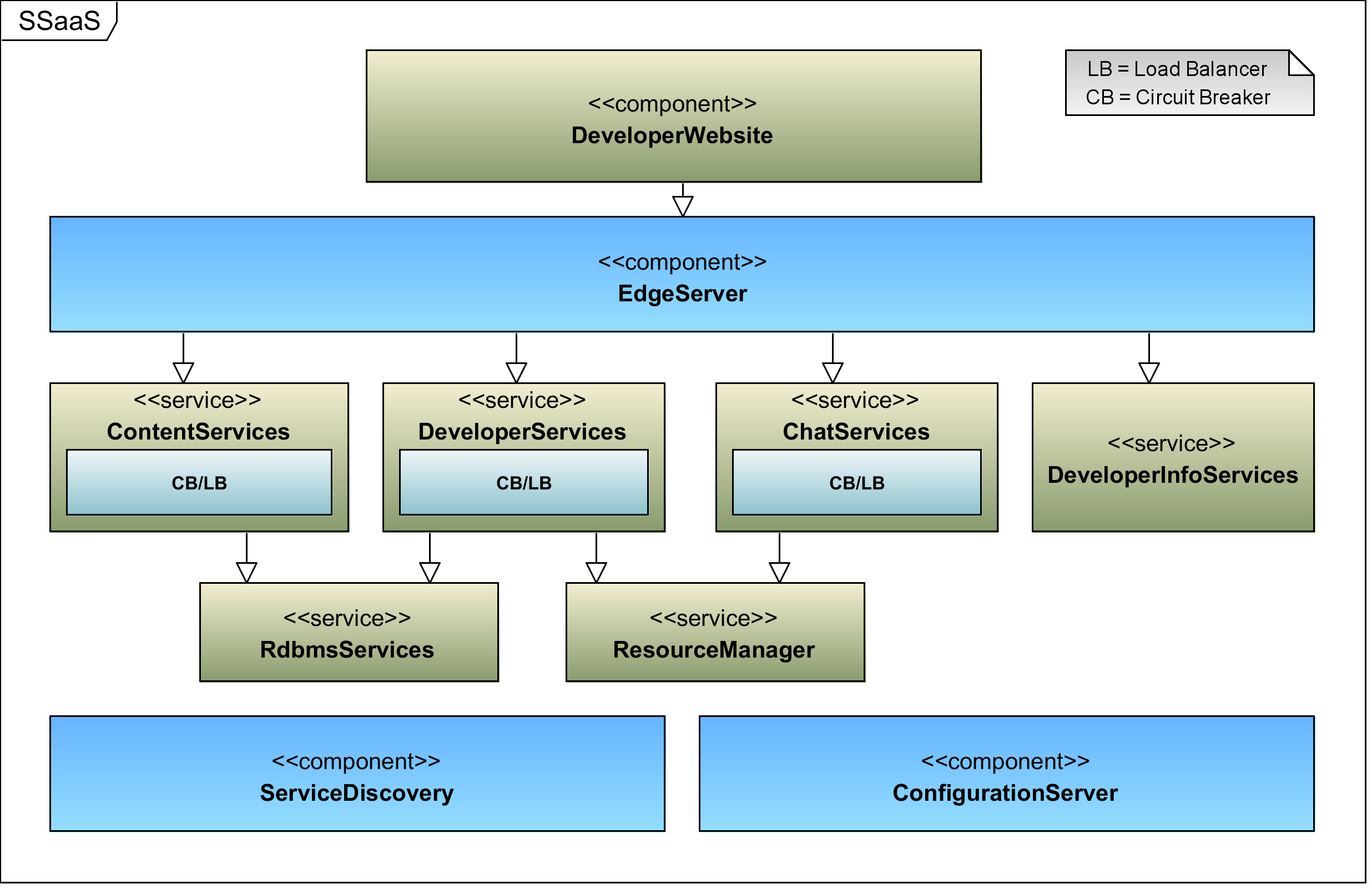}
  \caption{Target architecture of SSaaS after the migration}\label{fig:target_architecture}
\end{figure}

In order to realize microservices architecture and to satisfy our new requirements, we transformed the core architecture of our system to a target architecture by undergoing some architectural refactorings. These changes included introducing microservices-specific components as explained in \secref{background} and re-architecting the current system as will be discussed in this section. The final architecture is depicted in \figref{target_architecture}.

The new technology stack for the development was including the Spring Boot~\footnote{http://projects.spring.io/spring-boot} for its embedded application server, fast service initialization, using the operating system's environment variables for configuration, and the Spring Cloud~\footnote{http://projects.spring.io/spring-cloud} Context and the Config Server to separate the configuration from the source code as recommended by Continuous Delivery. Additionally, we chose the Netflix OSS~\footnote{http://netflix.github.io} for providing some of the microservices-specific components, i.e. Service Discovery, and the Spring Cloud Netflix that integrates the Spring framework with the Netflix OSS project. We also chose Eureka for Service Discovery, Ribbon as Load Balancer, Hystrix as Circuit Breaker and Zuul as Edge Server, that all are parts of the Netflix OSS project. We specifically chose Ribbon among other load balancers, i.e. HAProxy~\footnote{http://www.haproxy.org}, because of its integration with the Spring framework and other Netflix OSS projects, in particular Eureka. Additionally, it is an internal load balancer, so we do not need to deploy an external one.

\subsection{How Did We Re-Architect the System and Refactor the Data?}
In the state-of-the-art about microservices~\cite{newman2015buildingmicro,stine2015migratemicro}, \emph{Domain Driven Design}~\cite{evans2004domain,vernon2013implementing} and \emph{Bounded Context}~\cite{evans2004domain,vernon2013implementing} are introduced as common practices to transform the system's architecture into microservices. As we did not have a complex domain, we decided to re-architect the system based on domain entities in DeveloperData. We put every set of cohesive entities into a service, such that the only one which can create and update that entity would be that service. For example, only the ChatServices service could update or create the chat metadata entities. Other services can only have copies of the data that they do not own, e.g., for the purpose of caching. However, they should be careful about synchronization with the master data as their copy could be stale. With respect to this discussion, the list of architectural changes to reach the target architecture is the following:

\begin{itemize}
  \item Letting the ChatServices service handle its metadata by itself and not inside the DeveloperData.
  \item Introducing a new Resource Manager service in order to reserve resources like databases. The entities related to Oracle server instances will be moved from DeveloerData to this service.
  \item Introducing a new service to handle developer's information and its registered services.
  \item Transforming DeveloperData from a library to a service. Therefore, DeveloperServices and ContentServices have to be adapted such that they can make service calls to DeveloperData instead of method calls. Please note that the remaining data in DeveloperData are just RDBMS entities like Table and Column.
\end{itemize}

\section{Migration Steps} \label{sec:migration_steps}
Migrating the system towards the target architecture is not a one-step procedure and should be done incrementally and in several steps without affecting the end-users of the system. Furthermore, as the number of services is growing, we need a mechanism for automating the delivery process. In this section, we describe how we migrated SSaaS using the following eight steps:

\begin{figure}[t]
  \centering
  \includegraphics[width=150px]{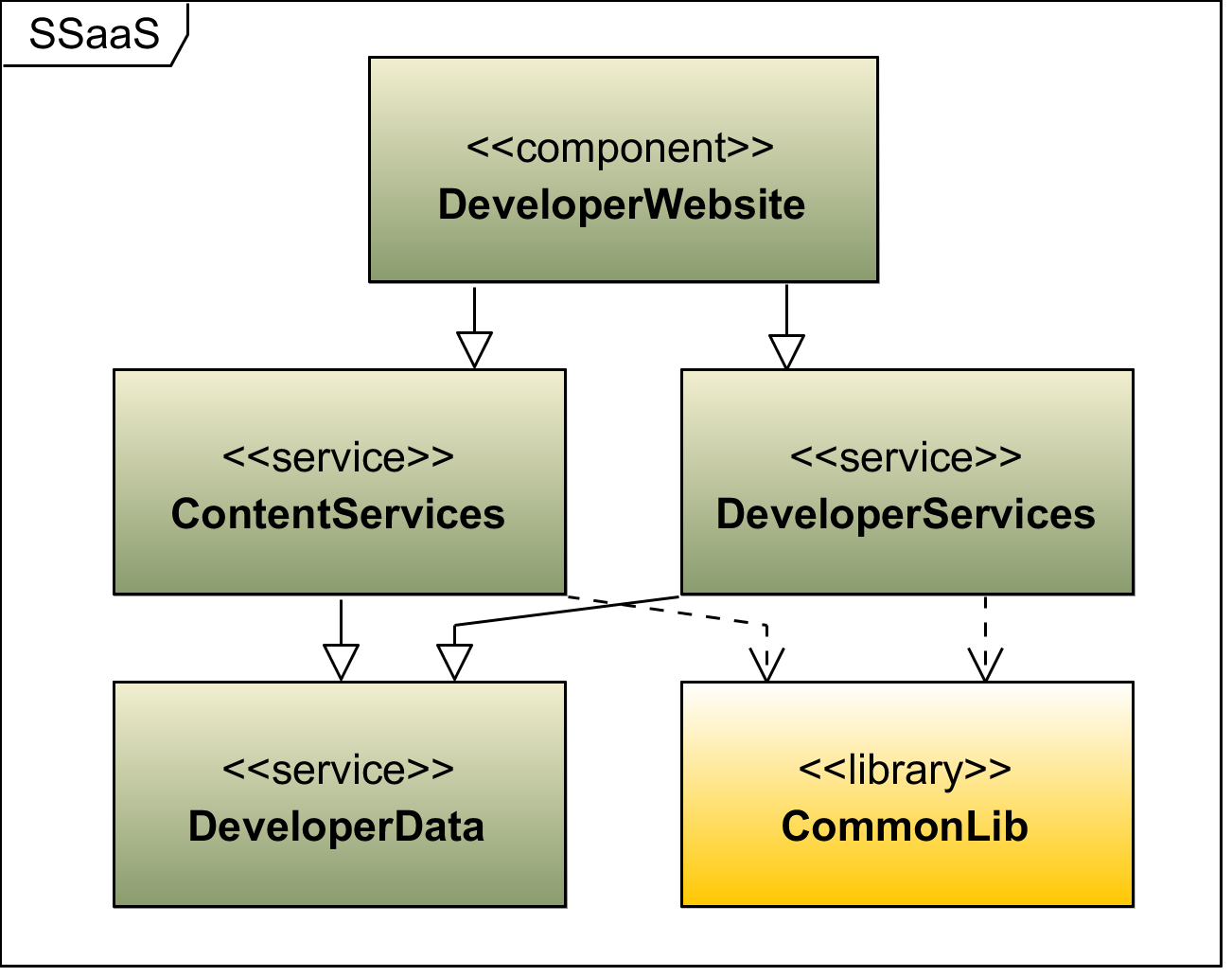}
  \caption{Transforming DeveloperData to a Service}\label{fig:step1}
\end{figure}

\subsection{Preparing the Continuous Integration Pipeline}
Continuous integration is the first step for having an effective Continuous Delivery. It allows developers to integrate their work with the others' early and often, and helps to prevent future conflicts~\cite{humble2010continuous}. To this end, a continuous integration (CI) server, an as-a-service or self-hosted code repository, and an artifact repository is needed. We chose Jenkins~\footnote{https://jenkins-ci.org} as the CI server, self-hosted Gitlab~\footnote{https://about.gitlab.com} as the code repository, and Artifactory~\footnote{http://www.jfrog.com/open-source} as the artifact repository (cf. \figref{final_pipeline}).

By adopting microservices, the number of services will increase. As each of these services can have a number of instances running, deploying them by virtualization is not cost-effective and can introduce a lot of computational overhead. Furthermore, we may need to use Configuration Management systems in order to create the exact test and production environments. \emph{Containerization} is a new trend that is well suited for microservices. By utilizing containers, we can deploy service instances with lower overheads than the virtualization, and in isolation. Additionally, we would not hear phrases like ``this works on my machine" anymore because we are using the exact environments and artifacts in both of the development and production environments. Another major benefit is \emph{portability} since we can deploy anywhere that supports containerization without any changes to our source codes or container images. Many public cloud providers such as Google and Amazon now have a support for containerization.

Docker~\footnote{https://www.docker.com} is a tool for containerization of applications, and it is now becoming the de-facto standard for containerization in industry. There is a pool of ready to use images in the Docker Hub, the central docker image repository, that can be pulled and customized based on specific needs. Docker Registry~\footnote{https://docs.docker.com/registry} is another project that let organizations to have a private docker image repository. As we are going to use Docker, we need Docker Registry to be in our pipeline as well.

To summarize, in this step, we installed and integrated the Gitlab, Jenkins, Artifactory and Docker Registry as a CI pipeline.

\subsection{Transforming DeveloperData to a Service}
In this step, we changed DeveloperData to use Spring Boot because of its advantages (see \secref{target_architecture}). Furthermore as shown in \figref{step1}, we changed it to expose its functionalities as a REST API. In this way, its dependent services would not be affected when the internal structure of DeveloperData changes. Since they have service-level dependency, the governance of DeveloperData entities will be done by a single service and DeveloperData would not act as an Integration Database~\cite{hohpe2004enterprise} for its dependent services anymore. Accordingly, we adapted DeveloperServices and ContentServices to use DeveloperData as a service and not as a Maven dependency.

\begin{figure}[t]
  \centering
  \includegraphics[width=150px]{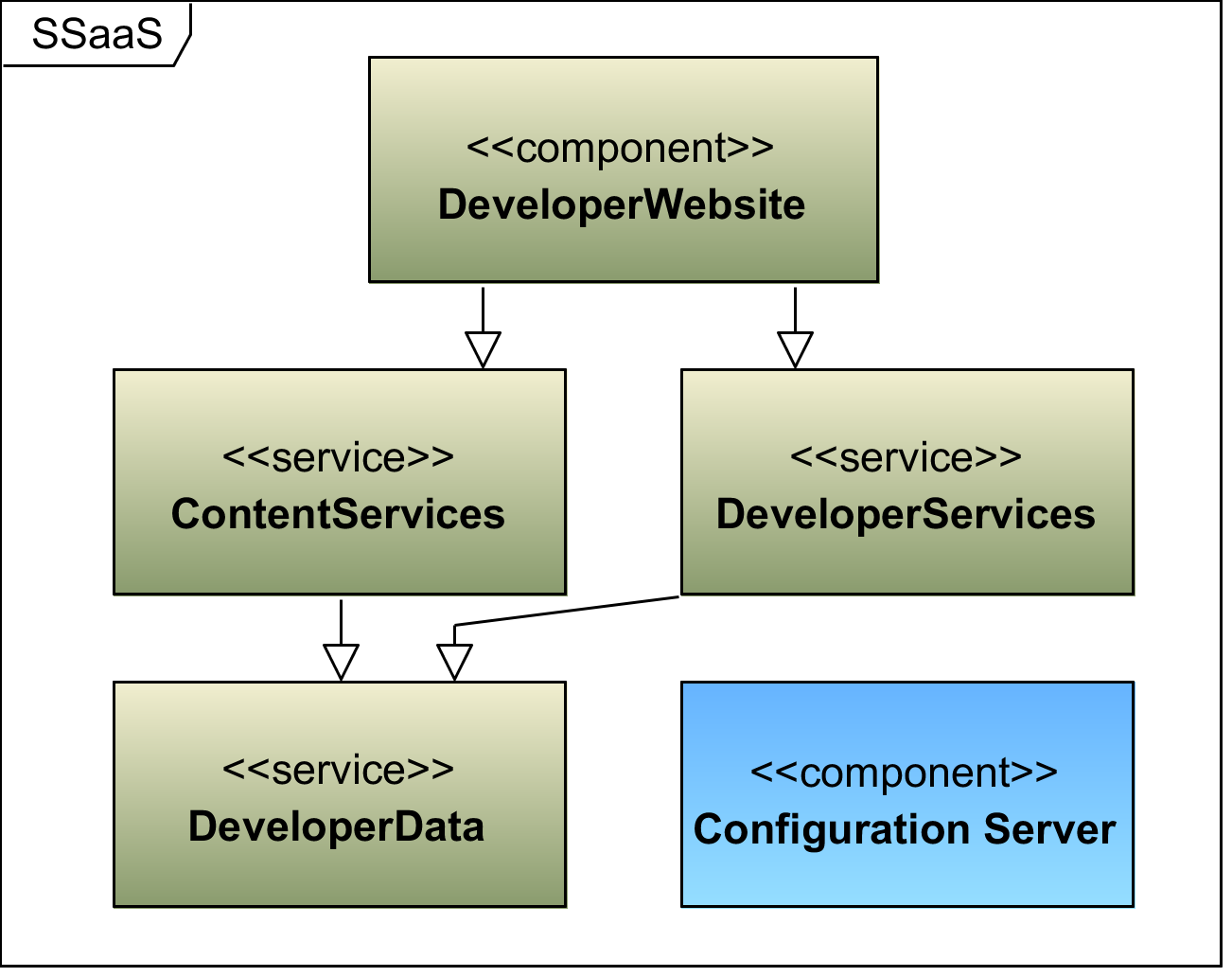}
  \caption{Introducing Configuration Server}\label{fig:step2}
\end{figure}

\subsection{Introducing Continuous Delivery}
A best practice in the Continuous Delivery is to separate the source code, the configuration, and the environment specification so that they can evolve independently~\cite{humble2010continuous}. In this way, we can change the configuration without redeploying the source code. By leveraging Docker, we removed the need for specifying environments since the Docker images produce the same behavior in different environments. In order to separate the source code and the configuration, we ported every service to Spring Boot and changed them to use the Spring Cloud Configuration Server and the Spring Cloud Context for resolving their configuration values (cf. \figref{step2}). In this step, we also separated services' code repositories to have a clearer change history and to separate the build life cycle of each service. We also created the Dockerfile for each service that is a configuration for creating Docker images for that service. After doing all of the mentioned tasks, we created a CI job per service and ran them in order to populate our repositories. Having the Docker image of each service in our private Docker registry, we were able to run the whole system with Docker Compose~\footnote{https://docs.docker.com/compose} using only one configuration file. Starting from this step, we had an automated deployment on a single server.

\subsection{Introducing Edge Server}
\begin{figure}[t]
  \centering
  \includegraphics[width=150px]{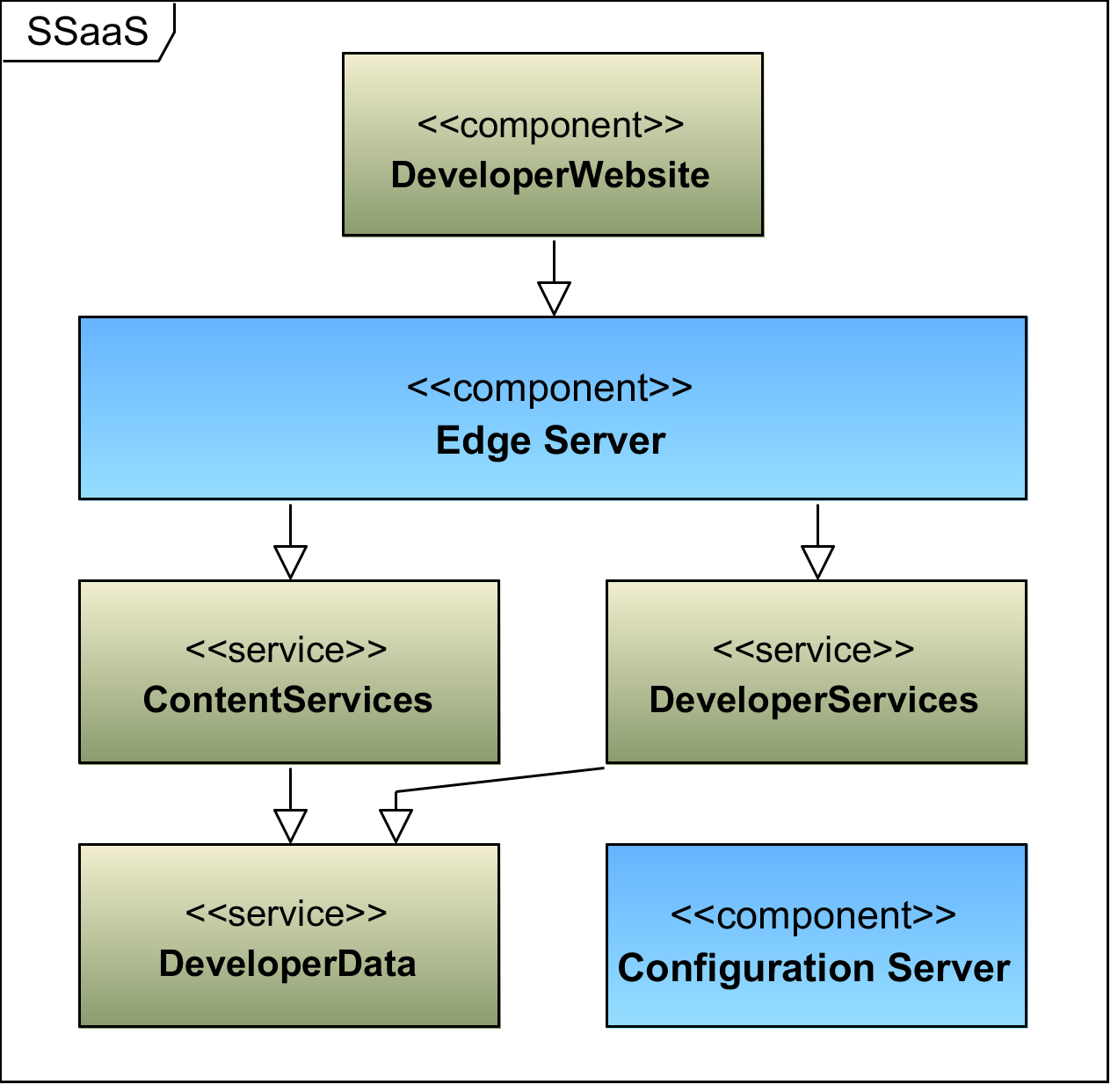}
  \caption{Introducing Edge Server}\label{fig:step3}
\end{figure}

As we were going to re-architect the system and it was supposed to change the internal service architecture, in this step, we introduced Edge Server to the system to minimize the impact of internal changes on end-users as shown in \figref{step3}. Accordingly, we adapted DeveloperWebsite.

\subsection{Introducing Dynamic Service Collaboration}
\begin{figure}[t]
  \centering
  \includegraphics[width=150px]{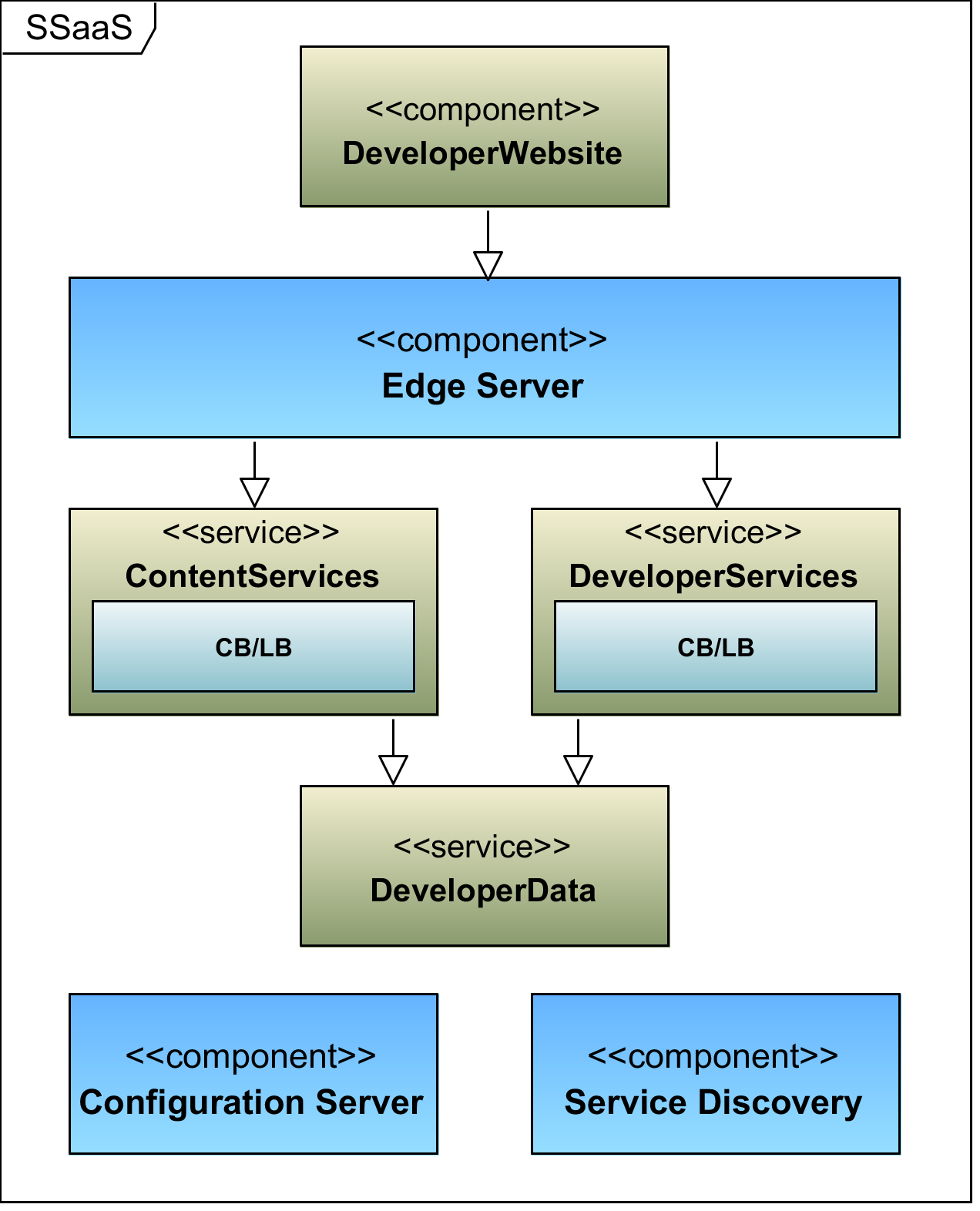}
  \caption{Introducing Dynamic Service Collaboration}\label{fig:step4}
\end{figure}

In this step, we introduced Service Discovery, Load Balancer and Circuit Breaker to the system as shown in \figref{step4}. Dependent services should locate each other via the Service Discovery and Load Balancer; and the Circuit Breaker will make our system more resilient during the service calls. By introducing these components to the system sooner, we made our developers more comfortable with these new concepts, and it increased our speed for the rest of the migration and of course, in introducing new services.

\subsection{Introducing Resource Manager}
\begin{figure}[t]
  \centering
  \includegraphics[width=150px]{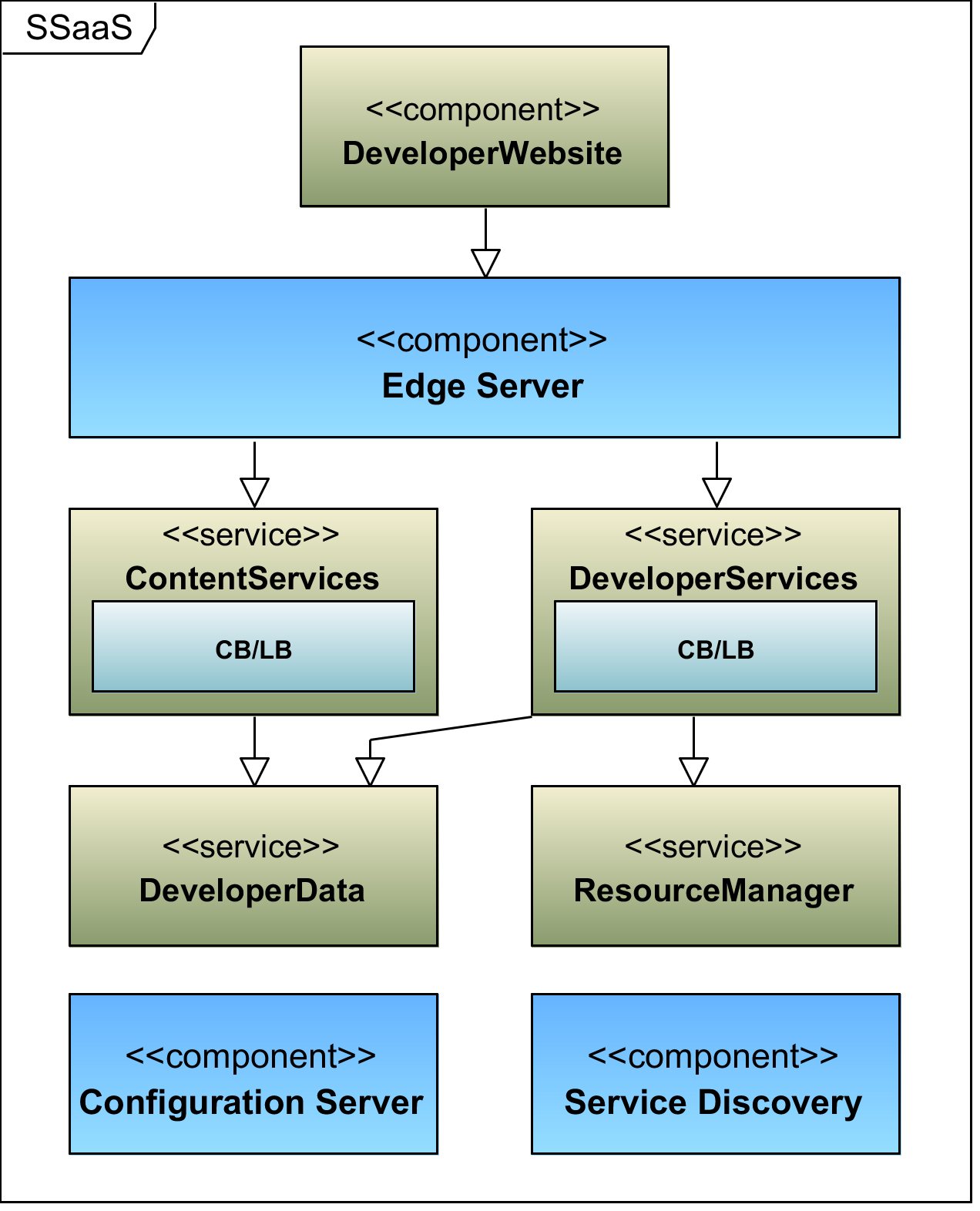}
  \caption{Introducing Resource Manager}\label{fig:step5}
\end{figure}

In this step, we introduced the Resource Manager by factoring out the entities that were related to servers, i.e. AvailableServer, from DeveloperData and introducing some new features, i.e. MySQL database reservation, for satisfying our chat service requirements (cf. \figref{step5}). Accordingly, we adapted DeveloperServices to use this service for database reservations.

\subsection{Introducing ChatServices and DeveloperInfoServices}
As the final step in re-architecting the system, we introduced the following services:

\begin{itemize}
  \item \emph{DeveloperInfoServices} by factoring out developer related entities (e.g., Developer) from DeveloperData.
  \item \emph{ChatServices} for persisting chat service instances metadata and handling chat service instance creations.
\end{itemize}

This led us to the target architecture as depicted in \figref{target_architecture}.

\subsection{Clusterization}
Compared to virtualization, one of the main features of containerization is its low overhead. Due to this feature, people started to make it more efficient by introducing lightweight operating systems, like CoreOS~\footnote{https://coreos.com} and Project Atomic, that only have the minimal parts to host many containers. Google Kubernetes~\footnote{http://kubernetes.io}, that has a good integration with the CoreOS, is a tool for easy deployments of containers on a cluster. Using Kubernetes, a container can be easily fetched from a private repository and deployed to a cluster with different policies. For example, a service can be deployed with three always available instances.

\begin{figure}[b]
  \centering
  \includegraphics[width=300px]{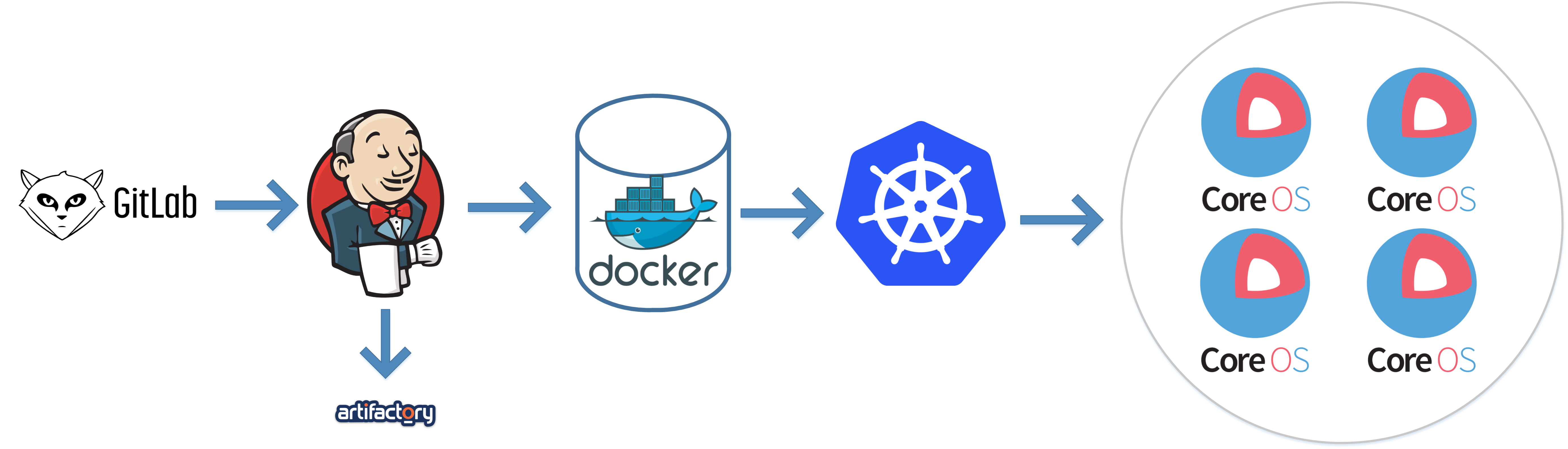}
  \caption{The final delivery pipeline}\label{fig:final_pipeline}
\end{figure}

In this step, we set up a cluster of CoreOS instances with Kubernetes agents installed on them. Next, we deployed our services on this cluster instead of a single server. The final delivery pipeline is shown in \figref{final_pipeline}.

In \secref{migration_steps}, we described the incremental process of migrating the SSaaS application towards the microservices architecture. This migration was actually performed in three dimensions: re-architecting the current system, introducing new supporting services, and enabling Continuous Delivery in the system. The important point to note is that how we incrementally evolved the system in all these three dimensions together. Despite the smoothness of the explained process, we faced several challenges in this process as well. \secref{lessons_learned} discusses some of the lessons we learned during this process.

\section{Lessons Learned} \label{sec:lessons_learned}
Migrating an on-premise application to a microservices architecture is a non-trivial task. During this migration, we faced several challenges that we were able to solve. In the following, we share some of the lessons we learned in this process that we think might be helpful for others who are also trying to migrate to microservices:

\begin{itemize}
  \item \emph{Deployment in the development environment is difficult:} Introducing new services to the system will put a big burden on developers. It is true that the application's code is now in isolated services. However, to run those services in their machines, developers need to deploy the dependent services as well. For example, the service registry should be deployed as well in order to have a working system. These kinds of deployment complexities are not normal for a novice developer. Hence, there should be a facility in place for setting up such a development environment with a minimum amount of effort. In our case, we chose the Docker Compose to easily deploy dependent services from our private Docker registry.

  \item \emph{Service contracts are double important:} Changing so many services that only expose their contracts to each other could be an error-prone task. Even a small change in the contracts can break a part of the system or even the system as a whole. Service versioning is a solution. Nonetheless, it could make the deployment procedure of each service even more complex. Therefore, people usually do not recommend service versioning in microservices. Thus, techniques like Tolerant Reader~\cite{daigneau2011service} are more advisable in order to avoid service versioning. Consumer-driven contracts~\cite{daigneau2011service} could be a great help in this regard, as the team responsible for the service can be confident that most of their consumers are satisfied with their service.

  \item \emph{Distributed system development needs skilled developers:} Microservices is a distributed architectural style. Furthermore, in order for it to be fully functional, it needs some supporting services like service registry, load balancer, and so on. Hence, to get the most out of microservices, those team members are needed who are familiar with these concepts and are comfortable with this type of programming.


  \item \emph{Creating service development templates is important:} Polyglot persistence and the usage of different programming languages are promises of microservices. Nevertheless, in practice, a radical interpretation of these promises could result in a chaos in the system and make it even unmaintainable. Consequently, having standards is a must in order to avoid chaos. Different languages and data stores can be used, but it should be in a controlled and standard way. As a solution, having service development templates for each leveraged language is essential. It would reduce the burden of development since people can easily fork the template and just start developing.


  \item \emph{Microservices is not a silver bullet:} Microservices was beneficial for us because we needed that amount of flexibility in our system, and that we had the Spring Cloud and Netflix OSS that made our migration and development a lot easier. However, as mentioned before, by adopting microservices so many complexities would be introduced to the system that require a lot of effort to be addressed. Therefore, these challenges should be considered before the adoption of microservices. In other words, maybe our problems could be solved more easily by applying another architectural style or solution.
\end{itemize}

\section{Conclusions and Future Work} \label{sec:conclusion}
In this paper, we explained our experience during the migration of an on-premise application to the microservices architectural style. In particular, we provided the architecture of our system before and after the migration and the steps that we followed for this migration. Furthermore, we highlighted the importance of Continuous Delivery in the process of adopting microservices. Finally, we discussed the lessons learned during this migration.

In future, we plan to consolidate these practices and develop a set of reusable patterns for migrating on-premise applications to microservices architectural style. These patterns should generalize the process that we used in this paper, but in a well-defined structure that can be instantiated independently, similarly to the approach that we devised in~\cite{jamshidicloud}.



\bibliographystyle{splncs}
\bibliography{refs}

\end{document}